\documentclass[11pt,british]{article}
\usepackage[utf8]{inputenc}
\newcommand{\B}{\boldsymbol}

\newcommand{\E}{\mbox{E}}

\newcommand{\CfD}{\mathrm{CfD}}
\newcommand{\FW}{\mathrm{FW}}
\newcommand{\SA}{\mathrm{SA}}
\newcommand{\SyS}{\mathrm{SS}}
\newcommand{\HY}{\mathrm{WA}}

\newcommand{\R}{\mathbb{R}}

\usepackage{tabularx}
\usepackage{setspace}
\usepackage{amsmath,amssymb}
\usepackage{natbib}
\usepackage{graphicx}
\usepackage[pdftex,colorlinks=true,pdfpagelabels,plainpages=false,linkcolor=black,citecolor=black,pagecolor=black,urlcolor=black,bookmarks]{hyperref}

\begin{document}

\title{Coping with area price risk in electricity markets: Forecasting Contracts for Difference in the Nordic power market}
\author{Egil Ferkingstad$^{*,1,2}$ and Anders L{\o}land$^2$ \\ 
$^1$Science Institute, University of Iceland\\Dunhaga 5, IS-107 Reykjavik, Iceland, and\\
$^2$Norwegian Computing Center\\ P. O. Box 114 Blindern, NO-0314 Oslo, Norway}

\date{\today}

\maketitle
\doublespacing
\begin{abstract}

Contracts for Difference (CfDs) are forwards on the spread between 
an area price and the system price. Together with the system price
forwards, these products are used to hedge the area price risk in the
Nordic electricity market. The CfDs are typically available for the
next two months, three quarters and three
years. This is fine, except that CfDs are not traded at
NASDAQ OMX Commodities for every Nord Pool Spot price area. We
therefore ask the hypothetical question: What would the CfD market price have
been, say in the price area NO2, if it had been traded? We build
regression models for each observable price area, and use Bayesian elicitation
techniques to obtain prior
information on how similar the different price areas are to forecast
the price in an area where CfDs are not traded. 

\smallskip
\noindent *Corresponding author. E-mail: egil@hi.is


\end{abstract}

\section{Introduction}

The Nordic electricity spot power market (Nord Pool Spot) is divided into
several price areas \citep{kristiansen04b,weron2006,benth08}, with the
system price being a common reference price. The different price areas 
result from capacity constraints. In theory, if an overall market
balance can be achieved without a need to utilise all available
capacity between neighbouring areas, the prices are equal in all areas. 

There is a parallel financial market, NASDAQ OMX Commodities, where
players in the market can hedge their positions through futures (days,
weeks) and forwards (months, quarters and years) against the system
price. However, nobody is exposed to the system spot price, but
rather to the area spot price. Therefore, 
the participants can in addition buy CfDs (Contracts for Difference) in order
to hedge the remaining difference between the system and price area
risk. The CfDs are typically available for the next two months, three quarters and three years. This is fine, except
that CfDs are not traded at NASDAQ OMX Commodities for every
Nord Pool Spot price area. There is still a need for 
hedging in those price areas where CfDs are not traded at the
exchange, but this is rather done through OTC (over-the-counter) trades. 
The price area risk is large, which is exemplified by the spot price in the NO1 (Oslo)
area. The NO1 spot price was quite close to the system spot price until
2007. During 2007, the NO1 price  fell below 20\% of the system
price. The price areas are defined by the transmission
system operators (TSOs). In Norway, the TSO Statnett has redefined the
price areas  five times between 2006 and 2011, making the
historic data not directly applicable for new price areas. 
We therefore ask the hypothetical question: What would the
CfD market price have been, say in today's NO2, if it had been traded?  

There are several reasons for being interested in this hypothetical
price. First and maybe foremost, we can then assess the
marking-to-mark value of possible CfDs entered OTC (over the counter)
for the non-traded areas. Second, we could possibly derive portfolios
of the traded CfDs to mimic a CfD in the non-traded areas. Third, CfD
prices can be used for internal risk and prognosis
considerations. Note that we seek to find the non-traded CfD prices,
which is fair in terms of the way the market operates, but may include 
(the non-observable) risk premiums. We will embroider this alternative
approach in the Discussion. 

This topic has not been discussed in the scientific literature, but Nord Pool Spot
system prices have been studied quite extensively 
\citep{benth08,erlewin10,botterud10}. There has been less focus
on Nord Pool Spot area prices, with some  noteworthy exceptions
\citep{kristiansen04b,fridolfsson09,jem_loland}.  Previous work on
Nordic CfD prices includes \cite{kristiansen04}, who 
 investigated hedging through these CfDs. He found that the contracts appear to be
 overpriced, but the results are preliminary due to a relatively new
 CfD market. \cite{marckhoff09} 
 found a significant relation between the spreads (Nord Pool Spot system minus
 area prices) and relative water
 reservoir levels for all areas except NO2 (Trondheim at the time) on more recent
 data (2001--2006). CfDs  are found to have significant risk
 premiums, with different signs and  magnitudes between areas. 

Our aim is to build a model for the expected CfD price   even for almost
brand new price areas where CfD products are not traded.  
We start by fitting standard regression models for the observed
CfDs for each price area and each CfD product. If the regression model
and its parameters had been the same for 
each observed price area, we could have validated it properly and used
it for every unobserved area as well with local covariate
values. Alas, the price areas are too different for this to
work. Alternatively, we could have found the unobserved CfD prices as
(model) averages of the observed ones, but the averaging weights can
not be found from the data at hand. Another approach would be to
estimate price area specific risk premiums in the spirit of
\cite{marckhoff09}. Since the risk premiums are not directly
observable and we are interested in the CfD prices, which are functions
of price expectations and risk premiums, modelling risk premiums would
be a detour here. We are therefore left with the fitted regression
models for each observed price area and the covariates for the
unobserved ones. The only viable path out of this deadlock is the use of Bayesian methods.  

We view this as a statistical elicitation problem
\citep{ohagan06}. Our method requires that we (or industry experts)
specify how much the (unobserved) price in 
area X resembles the (observed) price area Y regarding
covariate 1, 2, and so on. Therefore, the covariates have to be
relatively few,  readily interpretable and describe the observed
CfD prices well. To assess the corresponding uncertainty, we
ask how many months of data this opinion corresponds to. 

The problem at hand almost has no solution, but it still deserves our best
efforts. Proper validation is impossible, and we must simply rely on
our modelling approach. Yet, we can show that our method provides
sensible forecast results. Corresponding forecast problems with a missing
response in some data cells and subjective information can benefit
from our approach.

We proceed with presenting the data (Section \ref{sec:data}), and outline 
our approach in detail (Section
\ref{sec:methods}). The method is demonstrated  (Section
\ref{sec:res}), followed by a discussion of our approach
(Section \ref{sec:concl}).

\section{Data\label{sec:data}}

\begin{figure}[ht]
  \centering
  \includegraphics[width=1\linewidth]{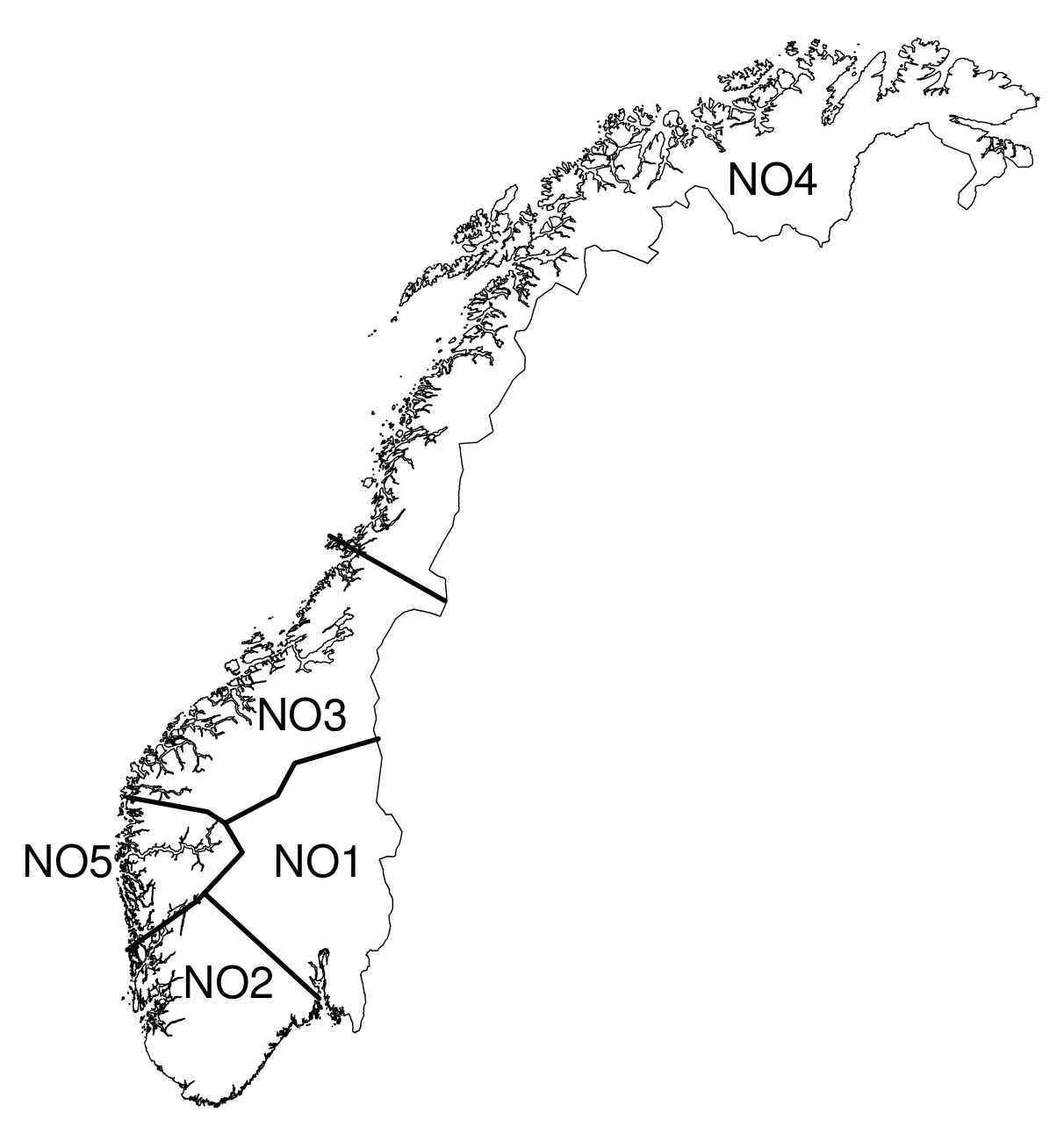}
  \caption{Current Norwegian price areas. Danish price areas DK1 and DK2 correspond to western and eastern Denmark, respectively, while Sweden and Finland each were single price areas during our data period.  \label{fig:priceareas}}
\end{figure}

\begin{figure}
   \centering
  \includegraphics[width=0.9\linewidth]{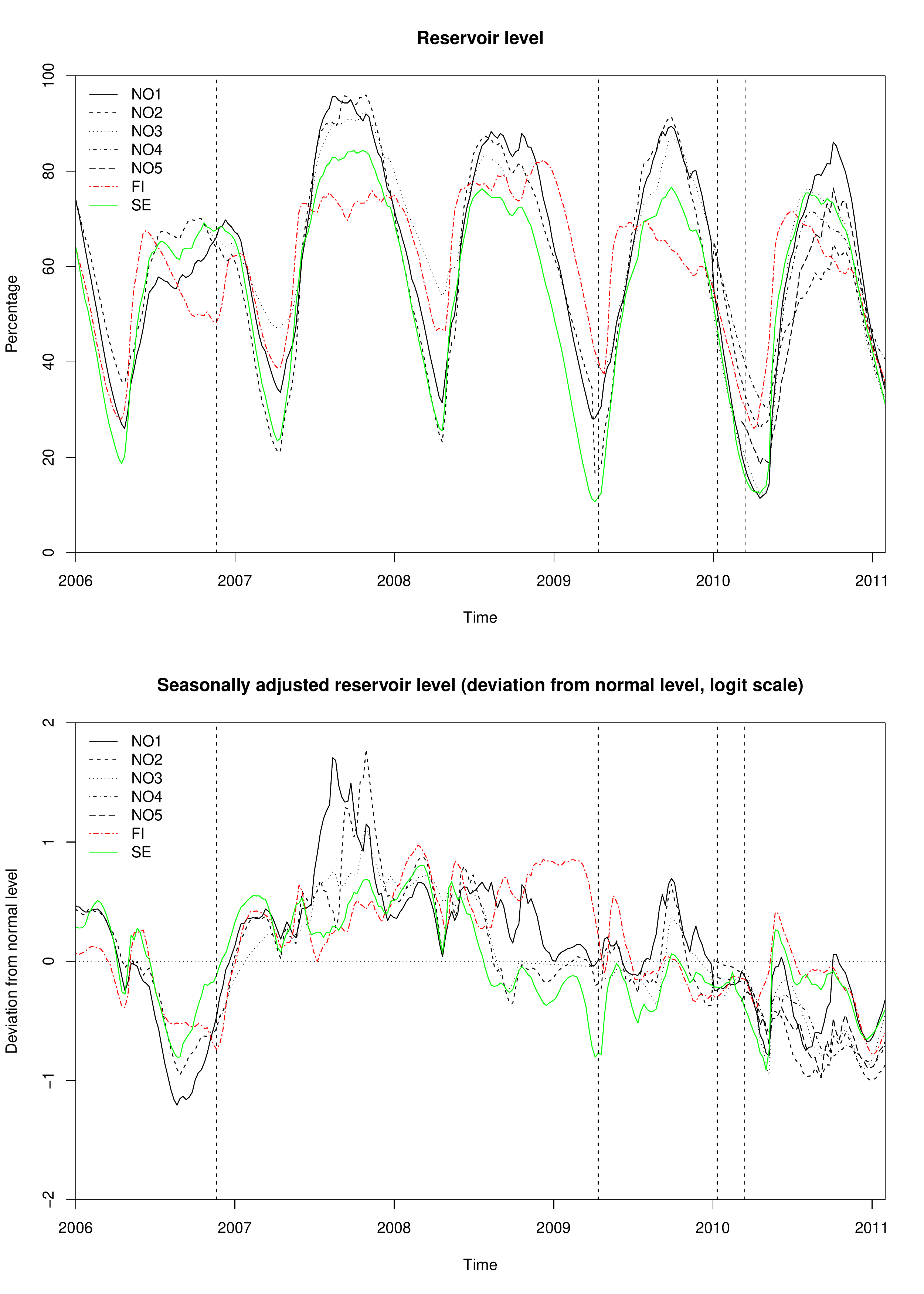}
  \caption{Upper: Water reservoir levels for the different price
    areas. 
    Lower: Seasonally adjusted water reservoir levels. Dashed vertical lines show dates when area definitions changed. \label{fig:water}}
\end{figure}

We use data from January 1st 2006 until January 31st 2011. During this
period, Statnett has redefined the Norwegian price areas five times
(marked by dashed vertical lines in Figures \ref{fig:water}--\ref{fig:predY1}),
while the Finnish (FI, one price area), Danish (DK1 and DK2, corresponding to western and eastern Denmark, respectively) and Swedish  
(SE) price area definitions have not changed. Svenska Kraftn\"at, the Swedish TSO, has
from the 1st of November 2011 divided Sweden into four areas, and
Norway is presently divided into five areas (NO1--NO5, Figure
\ref{fig:priceareas}). 

\begin{figure}[ht]
  \label{fig:spotprices}
  \centering
  \includegraphics[width=1\linewidth]{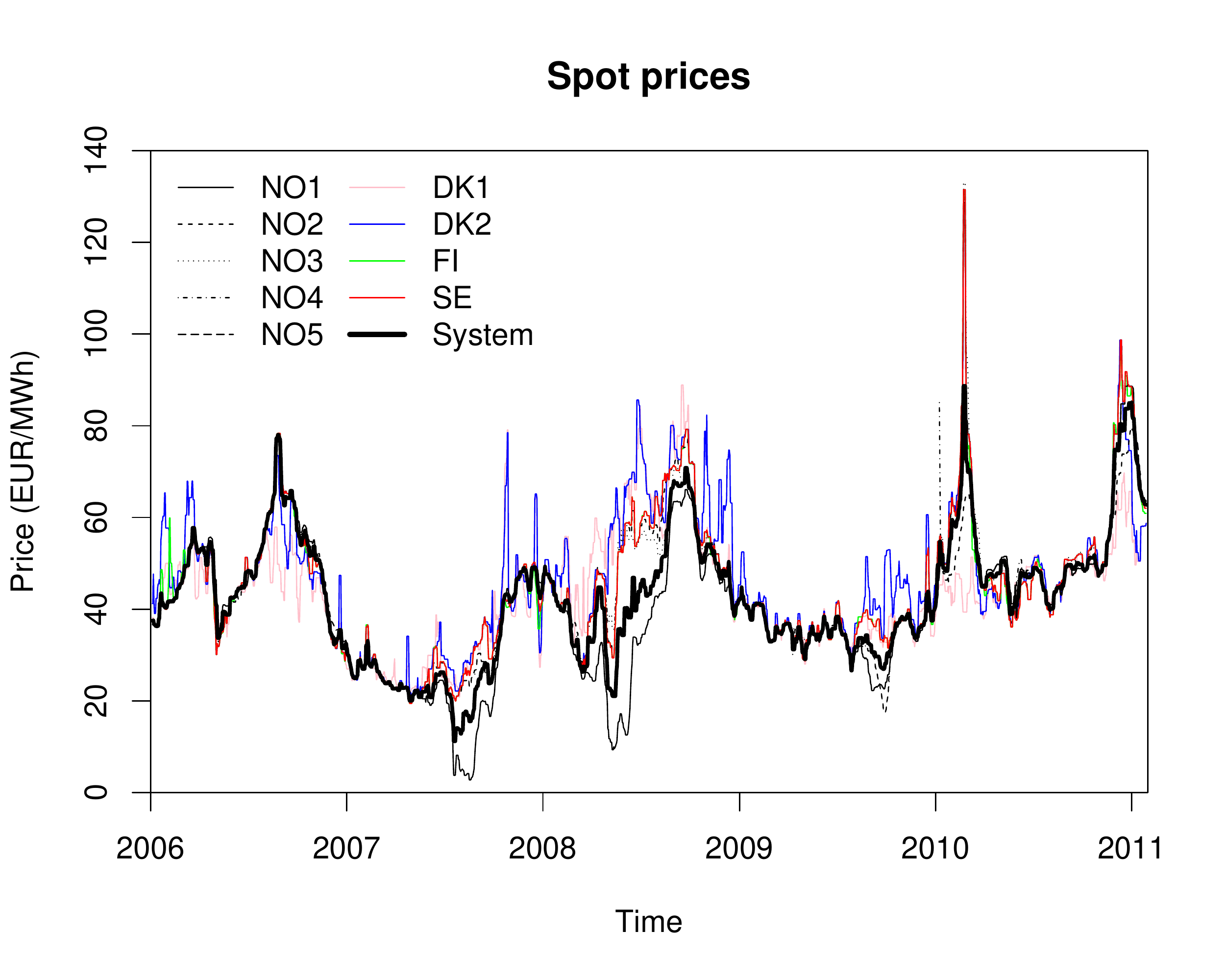}
  \caption{Area spot prices, with system spot price (heavy black curve)}
\end{figure}

We use the following data and nomenclature: $\CfD_{tah}$: CfD price for day
$t$, area $a$, horizon $h$, $\FW_{th}$: Forward price for day $t$,
horizon $h$, $\SA_{ta}$: Area spot price for day $t$, area $a$, $\SyS_t$:
System spot price for day $t$ and $\HY_{ta}$: Reservoir level (seasonally
adjusted, see below) for day $t$, area $a$.
Day $t=$ 1.1.2006, 2.1.2006, $\ldots$, 31.1.2011. The price areas $a$
include FI, DK1, DK2, SE, NO1--NO5.
Figure~\ref{fig:spotprices} shows the area and system spot prices.

 The horizon can be 1-2 months, 1-3 quarters or  1-3 years, which we 
write as $h \in \{\text{M1, M2, Q1, Q2, Q3, Y1, Y2, Y3}\}$.
Area spot prices are available for all areas, while CfDs are available
for all areas except NO2--NO5. The CfD and forward data are not strictly
daily: First, CfDs  and forwards are only sold on working
days. Second, and most important, CfDs for a particular area are not
actually sold each working day, even though a CfD price is recorded
-- i.e., if no CfD is sold on a working day $t$, the recorded price
is just the price at the most recent  $t' < t$ when a CfD was
sold. Thus, if we observe equal CfD prices for two subsequent days, it
may be either because 1) a CfD were actually sold at the same price
both days or 2) no CfD was sold on the second day. Unfortunately, we do
not have data on sales volume, but it is known that volumes can be rather low. It is, however, easy to incorporate volumes as weights in our model. We do not suspect that the general picture would be very different with the inclusion of volumes.

We may define the
\emph{area $a$ specific forward price} by  
\begin{align}\label{eq:cfd}
 \FW_{th} + \CfD_{tah}.
\end{align}
Since the area specific forward price must be positive, $\CfD_{tah} \in
(-\FW_{th},\infty)$, and we proceed with modelling $\CfD_{tah}$ on
the original scale.  

The covariates  are chosen because they in theory should
have predictive power and because they are readily
interpretable. \cite{kristiansen04} formulated a risk premium model
for the CfDs where the CfD is the discounted, expected difference
between the future area spot price and the future system spot
price. In a market without flexibility, storage capacity (like gas
storage or water reservoirs) and risk neutral players,  the current difference
between  the area and system spot price should be a good forecast for the future
CfD price. To relax this assumption somewhat, we include both the area
and the system spot price as covariates instead of the difference
between them. 

The CfD price can be seen as an absolute or relative deviation from
the system forward price \eqref{eq:cfd}. 
If the deviation is relative, the forward price
level should be of importance for the CfD, and we  therefore
include the corresponding  forward price (with the same
delivery period as for the CfD) as a covariate.

It is reasonable to assume that the volatility will increase for
products that are close to maturity \citep{aas04}. Time to maturity
was included as a covariate in a preliminary analysis, but turned out
to be non-significant. We believe the reason is that conceivable time
to maturity effects are included in $\FW_{th}$.

Since \cite{marckhoff09}  found a significant relation between the
empirical spreads and relative water reservoir levels for most
price areas, we include the reservoir level $\HY_{ta}$ as well for the
price areas with hydro power (Sweden, Finland and all Norwegian
ones, see Figure \ref{fig:water}). 
The series were seasonally adjusted by subtracting seasonal terms
\begin{align*}
\lambda_{ta} = \gamma_a^{(0)} + \sum_{j=1}^2{\gamma^{(1)}_{a,j} \sin \left( \frac{2 \pi
j t}{52} \right)
+ \gamma^{(2)}_{a,j} \cos \left( \frac{2 \pi j t}{52} \right), }
\end{align*}
which were estimated by least squares regression on logit-transformed
variables (to transform from $(0,100\%)\rightarrow \R$).
This was done in order to have variables that represent deviations
from a normal water reservoir level. The $\lambda_{ta}$ term was
estimated with data from 2006 to 2011 for every historical price area,
and the corresponding residuals for the applicable price area
definition were used.

\section{Methods\label{sec:methods}}

Consider a specific horizon, for example one quarter ahead.
Then, for each price area, we assume the linear regression model
\begin{equation}
\CfD_{ta}=\beta_{\FW}^{(a)}\FW_t+\beta_{\SA}^{(a)}\SA_{ta}+\beta_{\SyS}^{(a)}\SyS_t+\beta_{\HY}^{(a)}\HY_{ta}+\varepsilon_{ta}.
\label{eq:linmod}
\end{equation}
Note that there is no intercept term in the above model. This is
for two reasons: First, due to the nature of the CfD as the difference
of a hypothetical ``area-specific forward price'' and the actual
system forward price, it is reasonable to assume that the expected CfD
should be zero if all the covariates in~\eqref{eq:linmod} are
zero. Second, it is not clear how one could elicit intercept terms for
areas without observed CfD. 
There are obviously common
features between the horizons, but the effect of each covariate is
slightly different for each horizon. In addition, there is plenty of
data for each horizon, so some sort of shrinking between horizons to
reduce the number of parameters is not necessary. We therefore  consider each
horizon separately.  

For areas $a \in \{$DK1, DK2, FI, NO1, SE$\}$ we observe both
$\CfD_{ta}$ and the explanatory variables, so we could fit the linear
model using ordinary least squares. However, for the areas $a \in
\{$NO2, NO3, NO4, NO5$\}$ we only observe the explanatory
variables. Therefore, we obviously cannot fit the linear model for
these 
areas. To avoid this problem, it might seem natural to let the
regression coefficients for each explanatory variable be equal over
areas, so that we effectively have one large model over all
areas. Unfortunately, this model fits the observed areas very
poorly. Using separate linear models for each area fits the observed
data quite well.  

Because of all this missing information, we find it most sensible to use
some type of elicitation approach in order to take advantage of domain
experts' prior knowledge of how the areas differ. A relatively simple
way of doing this is to keep the linear model~\eqref{eq:linmod} for
the ``with-CfD'' areas, while assuming that the regression
coefficients for the ``without-CfD'' areas are given as a weighted
average of the coefficients for the ``with-CfD'' areas. The weights
are provided by the domain expert, together with an assessment of how
confident the expert is in the accuracy of the weights.

In more compact vector form, we may write this model as
\[
\B\CfD_a=\B X_a\B \beta_a+\B \varepsilon_a,
\]
where $\B\CfD_a$ denotes the vector of CfDs at all time points
for area $a$, $\B X_a$ is the matrix of covariates, $\B \beta_a$ the
vector of regression coefficients, and $\B \varepsilon_a$ is a vector of noise terms. 
From now on, for ease of notation, we number the areas, and order them
such that $a=1,\ldots,q$ are the ``with-CfD'' areas, while
$a=q+1,q+2,\ldots,m$ are the ``without-CfD'' areas.

We further assume that
\[
\B\CfD_a|\B \beta_a,\sigma_a^2,\B X_a \sim N(\B X_a\B \beta_a, \sigma_a^2 I)
\]
with a uniform prior on $(\B \beta_a, \log \sigma_a)$, i.e.
\begin{equation}
\label{eq:linmodprior}
p(\B \beta_a,\sigma_a^2|\B X_a) \propto \sigma_a^{-2}
\end{equation}
(which is a standard non-informative prior for linear regression).

For the unobserved areas $a>q$ we also assume that
\[
\B\CfD_a |\B \beta_a,\sigma_a^2,\B X_a \sim N(\B X_a\B \beta_a, \sigma_a^2 I),
\]
but here we cannot estimate the regression parameters directly, since we do not have any observed CfD.
We assume that each regression parameter for the unobserved areas $a>q$ is a weighted average of the regression coefficients for the observed areas, so that
\[\beta_{ia} = \sum_{j=1}^q w_{ij} \beta_{ij}, \ a>q,\]
where $\sum_{j=1}^q w_{ij}=1$ for each $i=1,\ldots,p$. Further, we assume that each vector $\B w_j=(w_{1j},\ldots,w_{qj})$ is Dirichlet distributed:
\[\B w_j \sim \text{Dirichlet}(\rho_{1j},\ldots,\rho_{qj},n).\]
The Dirichlet distribution was chosen because it gives a simple, interpretable way (described below) for the expert to quantify her level of certainty in her prior judgments. Also, viable alternatives to the Dirichlet prior seems to be lacking in this case~\citep[Section 6.5]{ohagan06}

The parameters $\rho_{1j},\ldots,\rho_{qj}$ and $n$ are determined
subjectively, using a structured (guided) elicitation procedure. Each
$\rho_{ij}$ may be interpreted as a measure of how much the effect of
covariate $j$ on the CfD price in area $a$ resembles the effect of
covariate $j$ on area $i$, relative to the other observed areas. While
this certainly sounds quite technical when expressed in such general
terms, it still remains amenable to elicitation. For example, we may
pose the question to an expert in the following way: ``Consider the
effect of hydrological balance (in NO2) on (the hypothetical) CfD
price in area NO2. How similar is this to the effect of the
hydrological balance (in NO1) on
the (observed) CfD price in NO1?'' By repeating this question for each
observed area, and scaling the answers so that they sum to one, we
obtain the parameters $\rho_{ij}$ for area NO2. We also want an
assessment of the level of certainty the expert is willing to attach
to her judgments. This is provided by the parameter $n$, which may be
elicited by the following question: ``Consider your answer to the
previous question. This is a question which also could  have been
answered using empirical data, if these were available. How many
months of data do you feel would give an information content
equivalent to your subjective assessment?''.

Suppose you are determining the vectors $\B w_j$ for NO2. The simplest
solution is to say that it will react to the covariates as in 
NO1. If you believe that the CfD price should react quicker to the
spot price signal, more weight could be put on the area and system
spot price ($SA_{ta}$ and $SS_{t}$) for the Danish areas DK1 and
DK2. Naturally, Danish hydro power is very limited, so the
corresponding reservoir content regression coefficients are set to zero. An
example of specified values for the parameters
$\rho_{1j},\ldots,\rho_{qj}$ for NO2 is given in Table
\ref{tab:elicitation}. 

\begin{table}[!h]
\begin{center}
\begin{tabular}{c|cccccc}
\textbf{NO2}  & DK1 & DK2 & FI & NO1 & SE & Months of data\\
\hline
$FW$ & 5\% & 5\% & 5\% & 75\% & 10\% & 1 \\
$SA$ & 5\% & 5\% & 5\% & 80\% & 5\% & 1 \\
$SS$ & 5\% & 5\% & 5\% & 80\% & 5\% & 1 \\
$HY$ & 0\% & 0\% & 5\% & 85\% & 10\% & 1 \\
\end{tabular}
\end{center}
\caption{Specified prior distribution for the parameters in the
  Dirichlet distribution for NO2. 
\label{tab:elicitation}}
\end{table}

Most weight is put on the NO1 area, since we
expect it to resemble NO2 the most. Except for the
missing Danish water reservoirs, 
all parameters are set to at least 5\%, since we are quite
uncertain. For the same reason, we feel that one month of data would
give informational content equivalent to our subjective assessment. We have experimented with different prior weights, and moderate changes does not affect the results too much.

After finishing the elicitation, the full model is fitted
using the following  Monte Carlo algorithm:
\begin{enumerate}
\item For each observed (``with-CfD'') area $i=1,\ldots,q$ and each covariate $j=1,\ldots,p$:
  \begin{enumerate}
  \item Draw $p_{ij}$
    from the Dirichlet distribution with parameters $\rho_{ij}$ and
    $n$.
  \item Draw regression parameters $\beta_{ij}$ from the posteriors of the linear model parameters for
    the observed areas (we used the function \texttt{sim} in the R package
    \texttt{arm}~\citep{arm}, see also Section~7.2 of~\citet{gelman2006}).
  \end{enumerate}
\item For each unobserved (``without-CfD'') area $i=q+1,\ldots,m$ and each covariate
  $j=1,\ldots,p$, calculate the predicted mean CfD $\mu_{it}=\sum_{j=1}^p \tilde
  \beta_{ij}x_{ijt}$, where $\tilde
  \beta_{ij}=\sum_{k=1}^q p_{kj} \beta_{kj}$.
\end{enumerate}
Repeating steps 1 and 2 above $N$  times then provides $N$ samples of
the predicted mean CfD $\mu_{it}$, which can be used for probabilistic
forecasting, for example using quantiles or the mean of the sampled
$\mu_{it}$.

\section{Results\label{sec:res}}

\begin{figure}[htn]
  \centering
  \begin{minipage}{0.45\linewidth}
    \includegraphics[width=\linewidth]{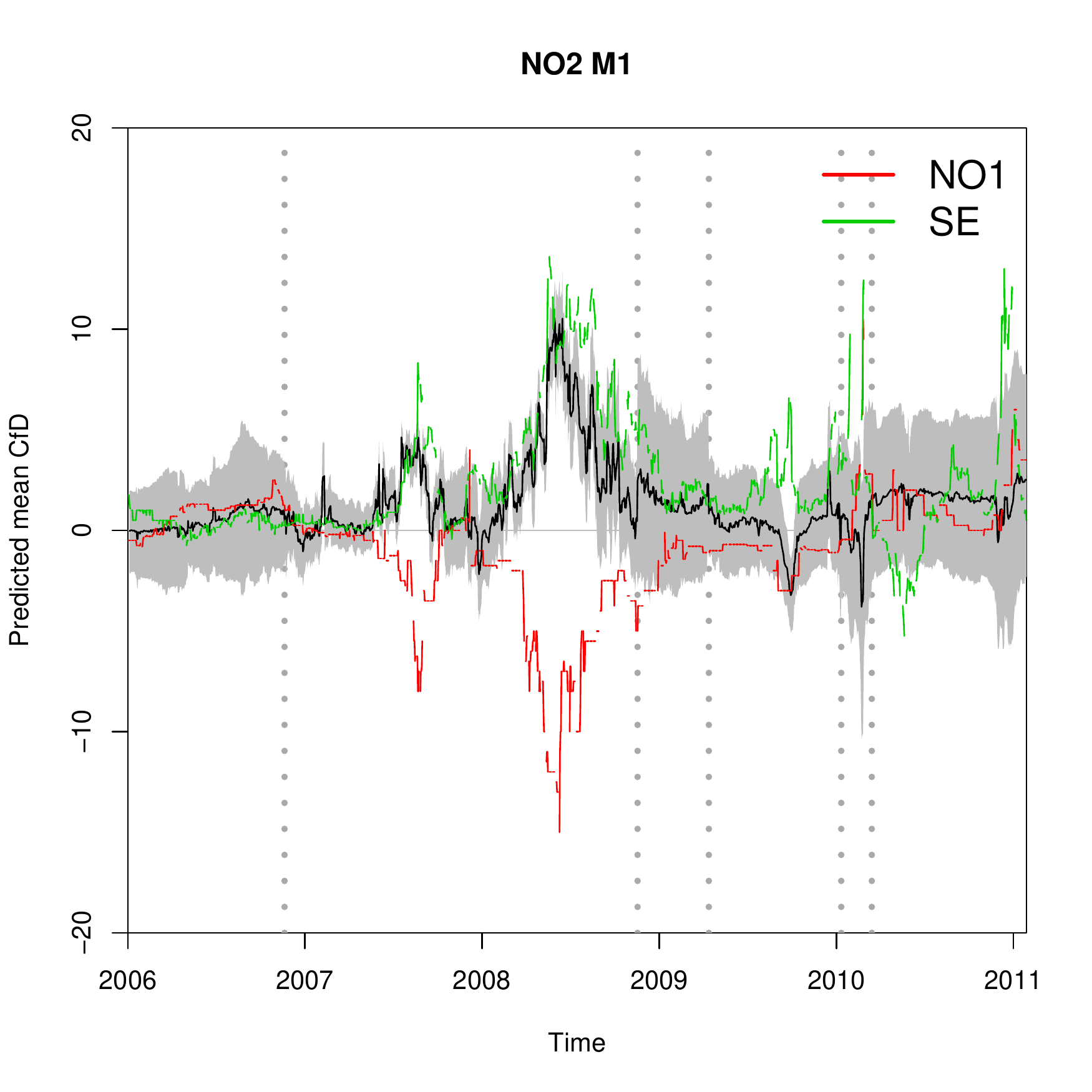}
    \includegraphics[width=\linewidth]{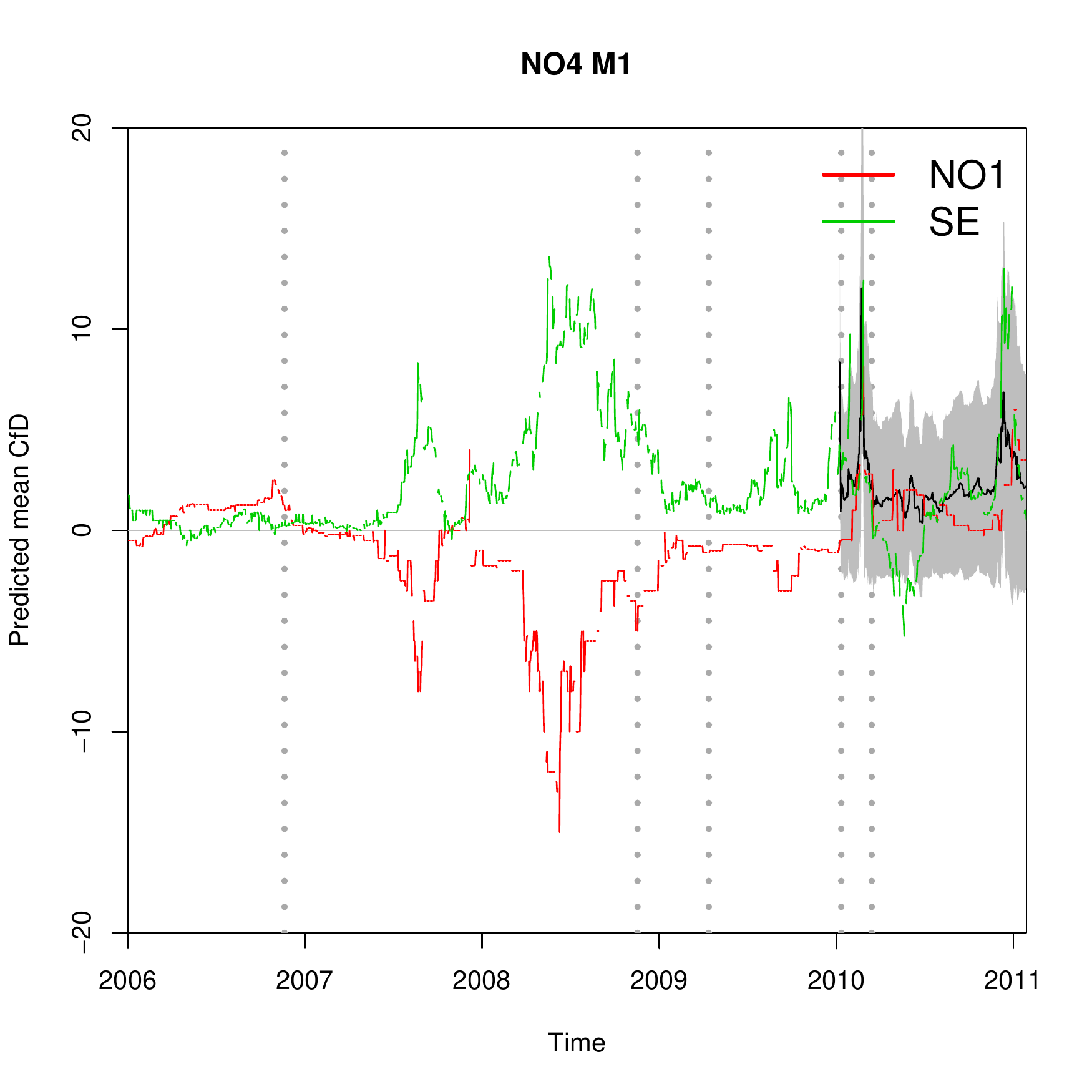}
  \end{minipage}
\begin{minipage}{0.45\linewidth}
    \includegraphics[width=\linewidth]{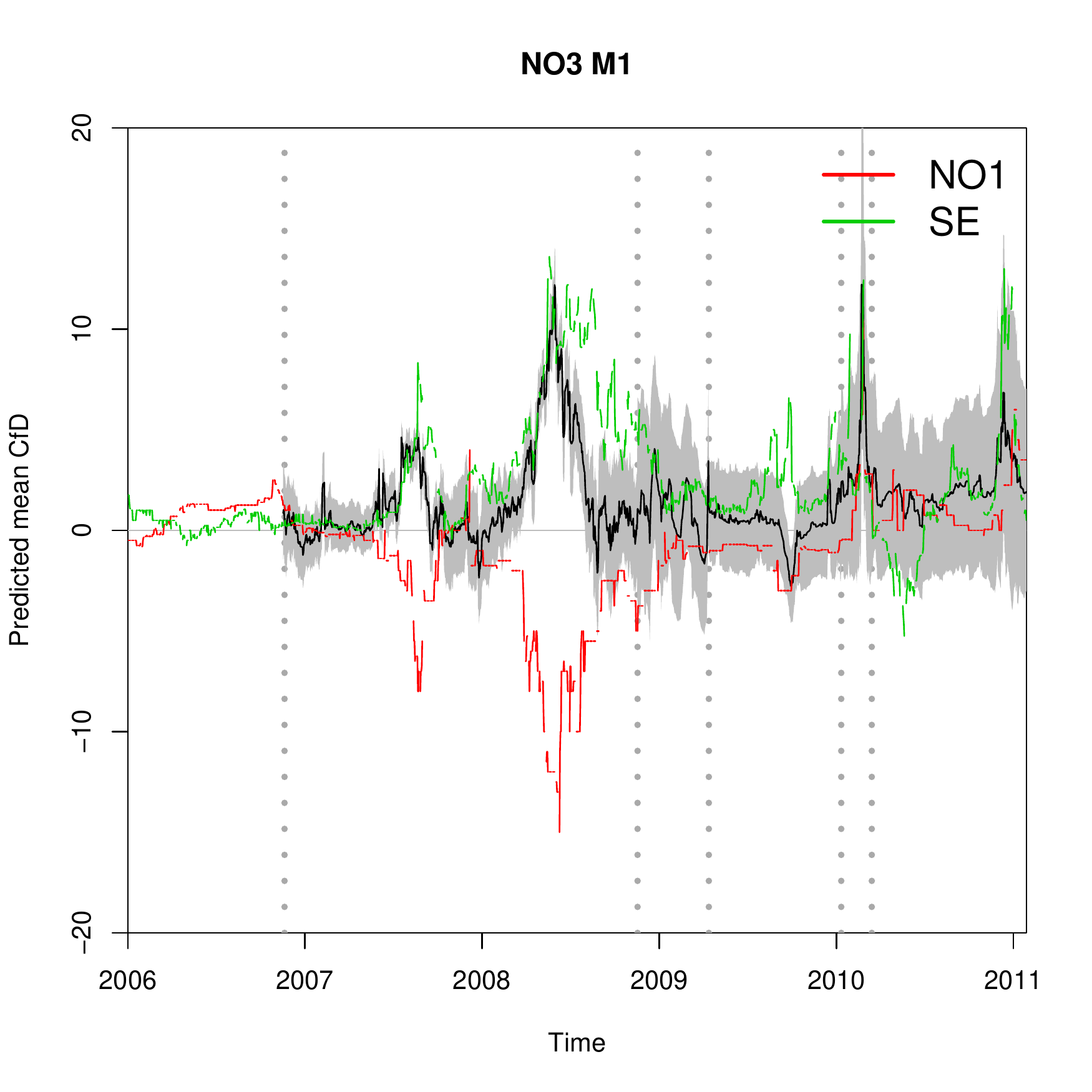}
    \includegraphics[width=\linewidth]{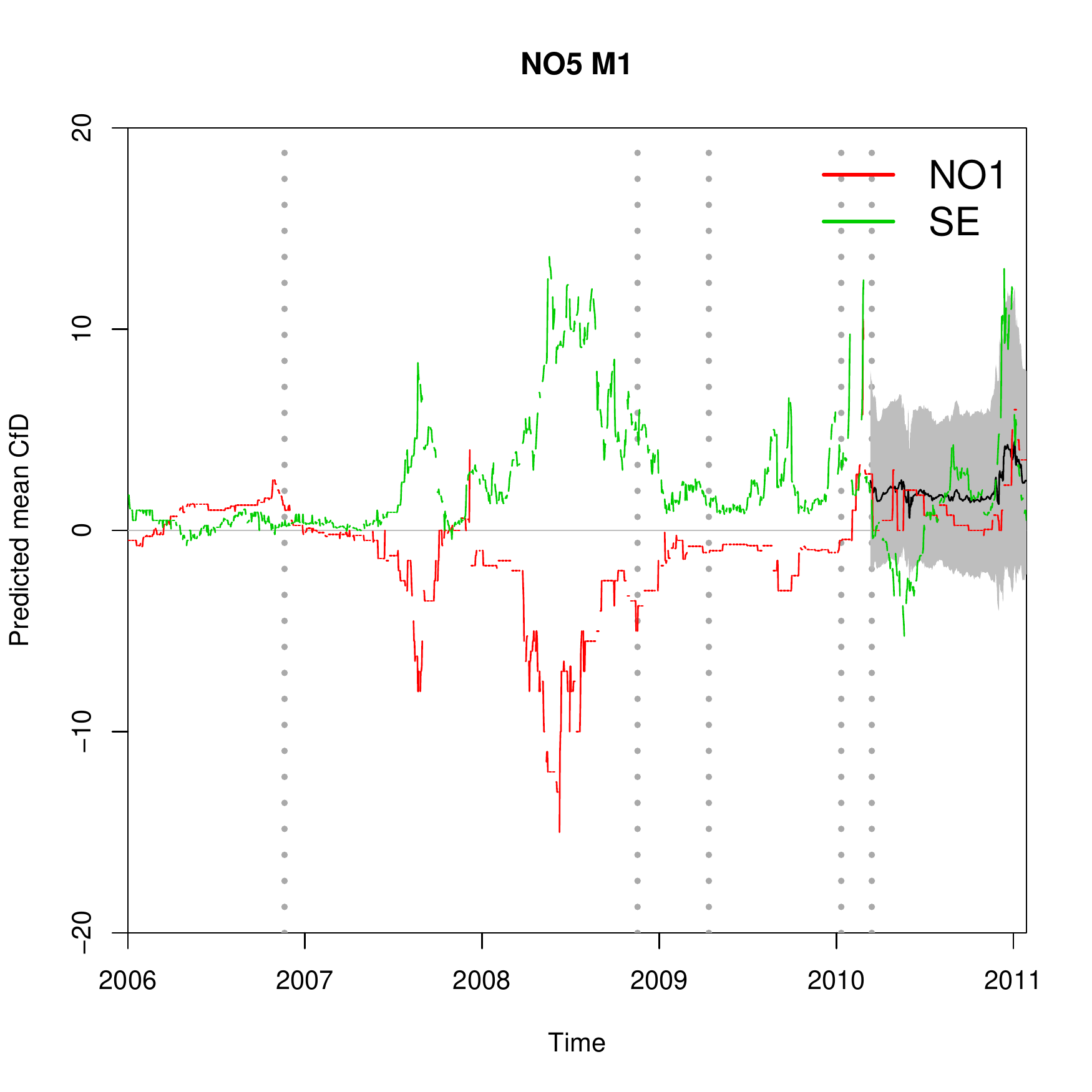}
  \end{minipage}
  \caption{Predicted CfDs for the M1 horizon.  A a 95\% prediction
interval is shown in grey. Dashed vertical lines indicate dates when the area
definitions changed.}
  \label{fig:predM1}
\end{figure}

\begin{figure}[htn]
  \centering
  \begin{minipage}{0.45\linewidth}
    \includegraphics[width=\linewidth]{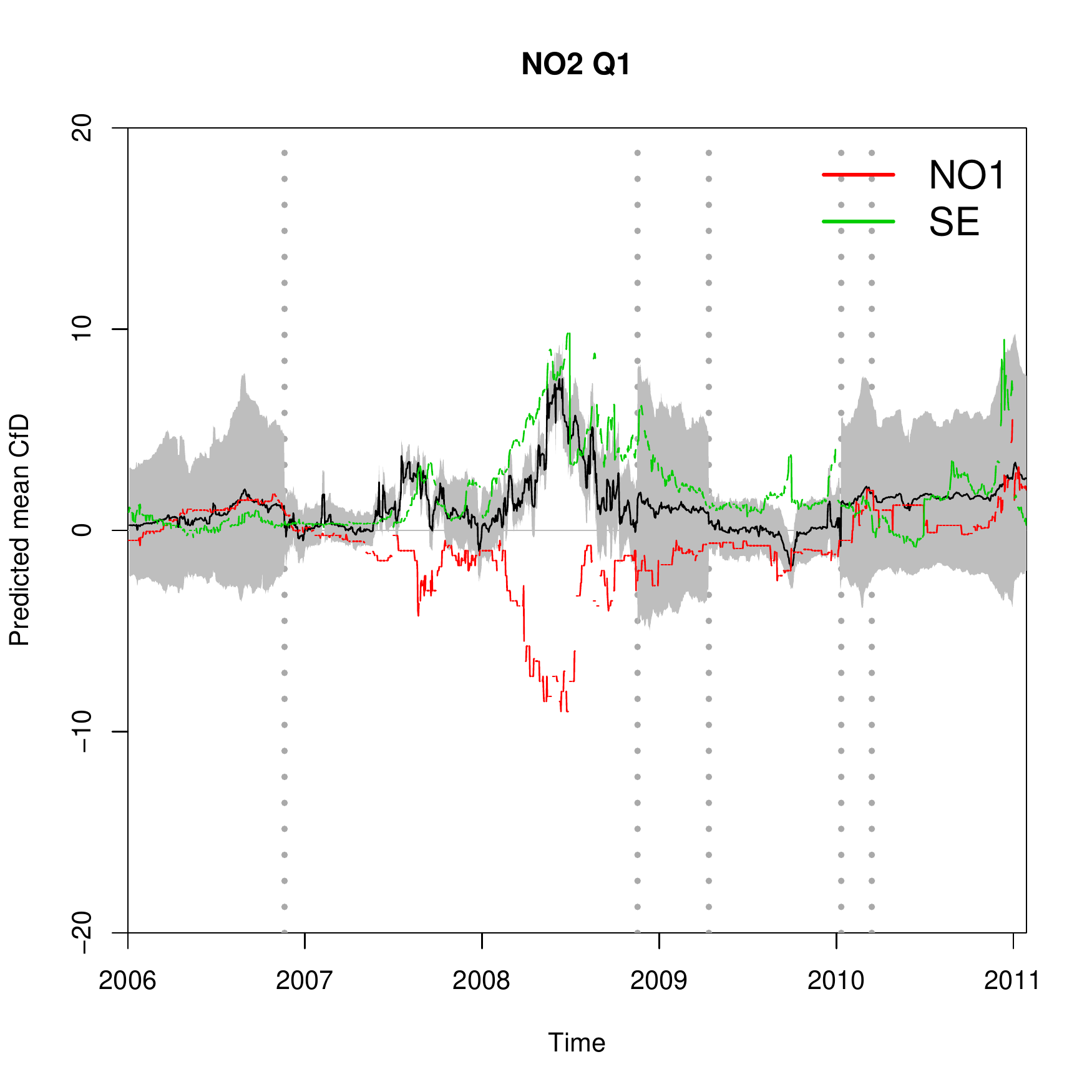}
    \includegraphics[width=\linewidth]{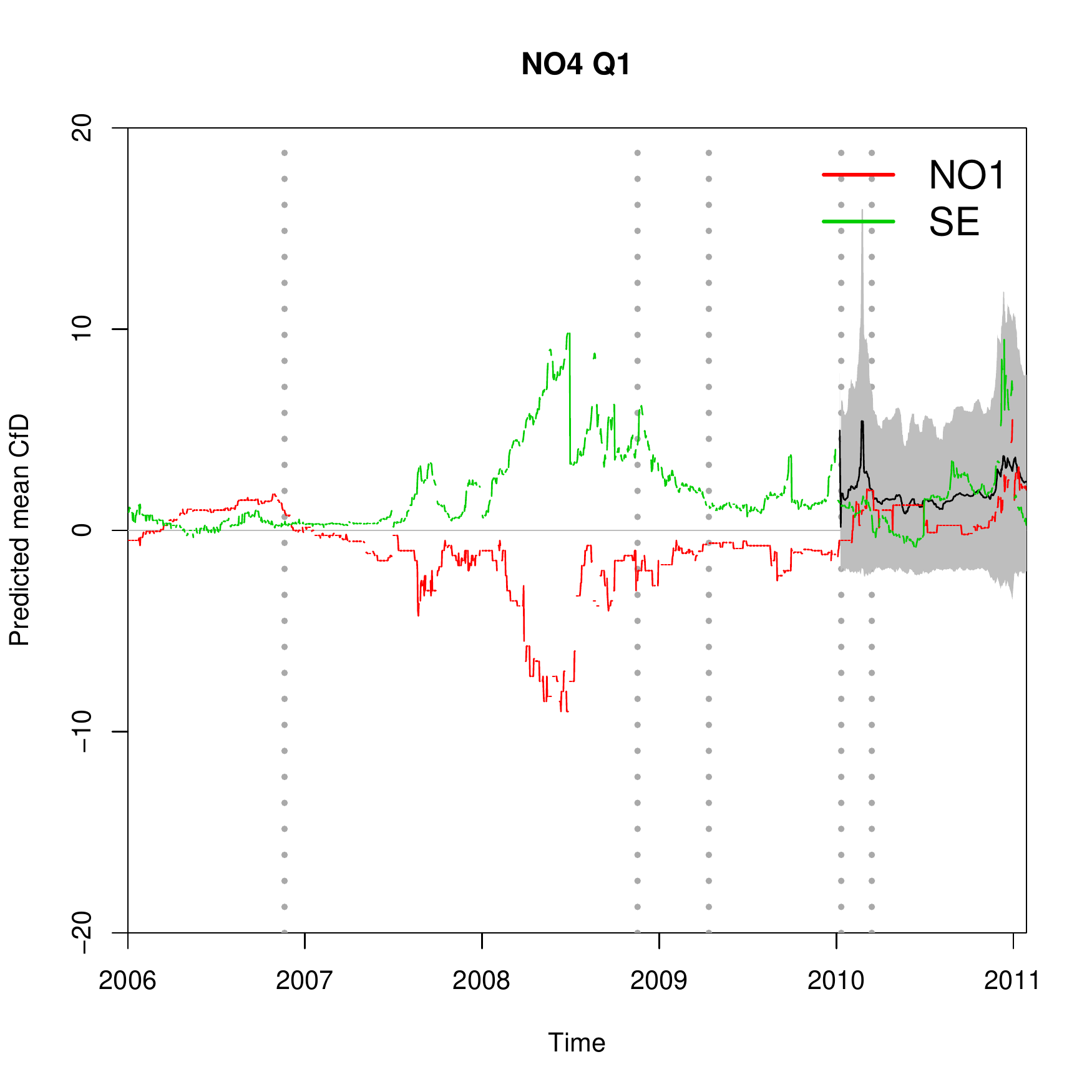}
  \end{minipage}
\begin{minipage}{0.45\linewidth}
    \includegraphics[width=\linewidth]{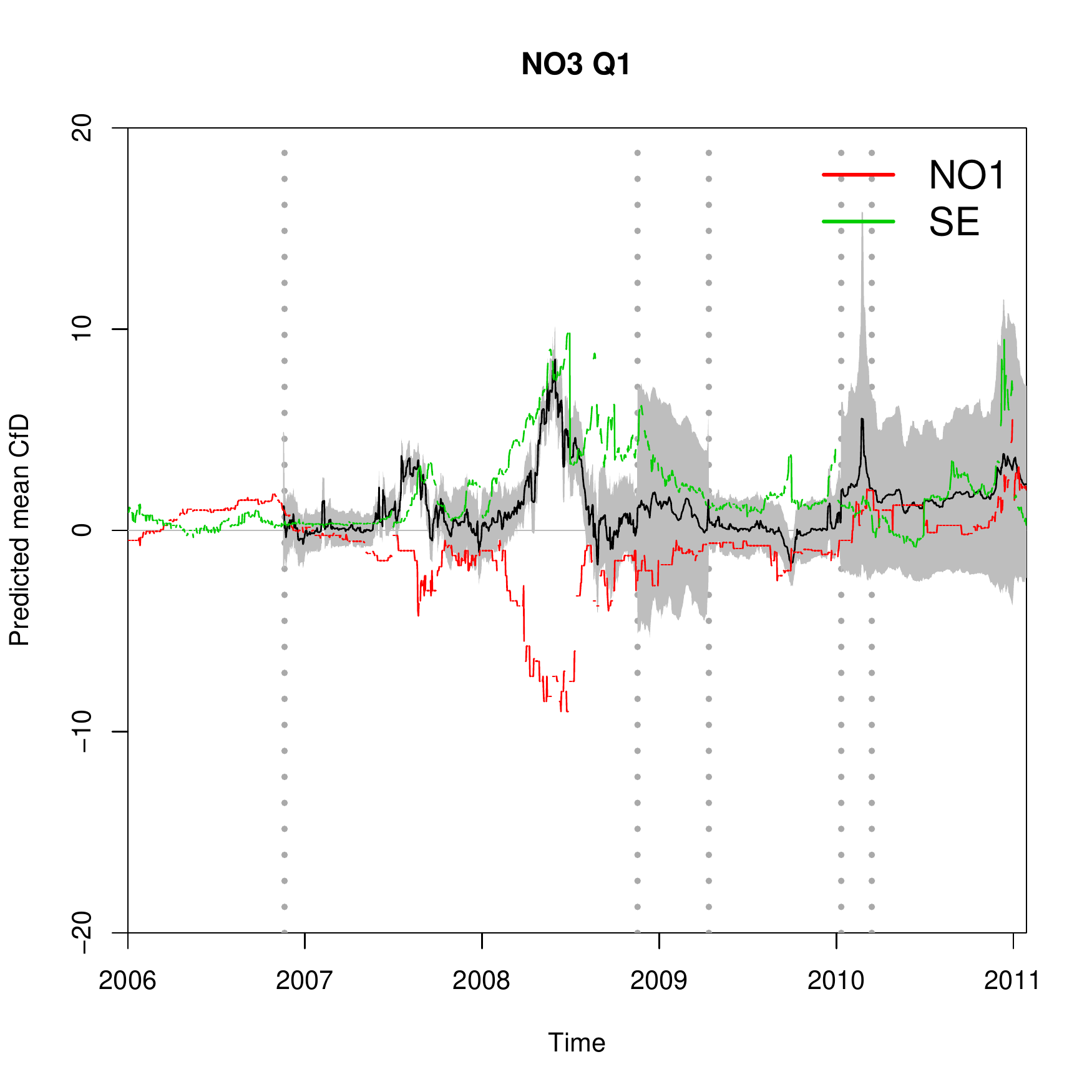}
    \includegraphics[width=\linewidth]{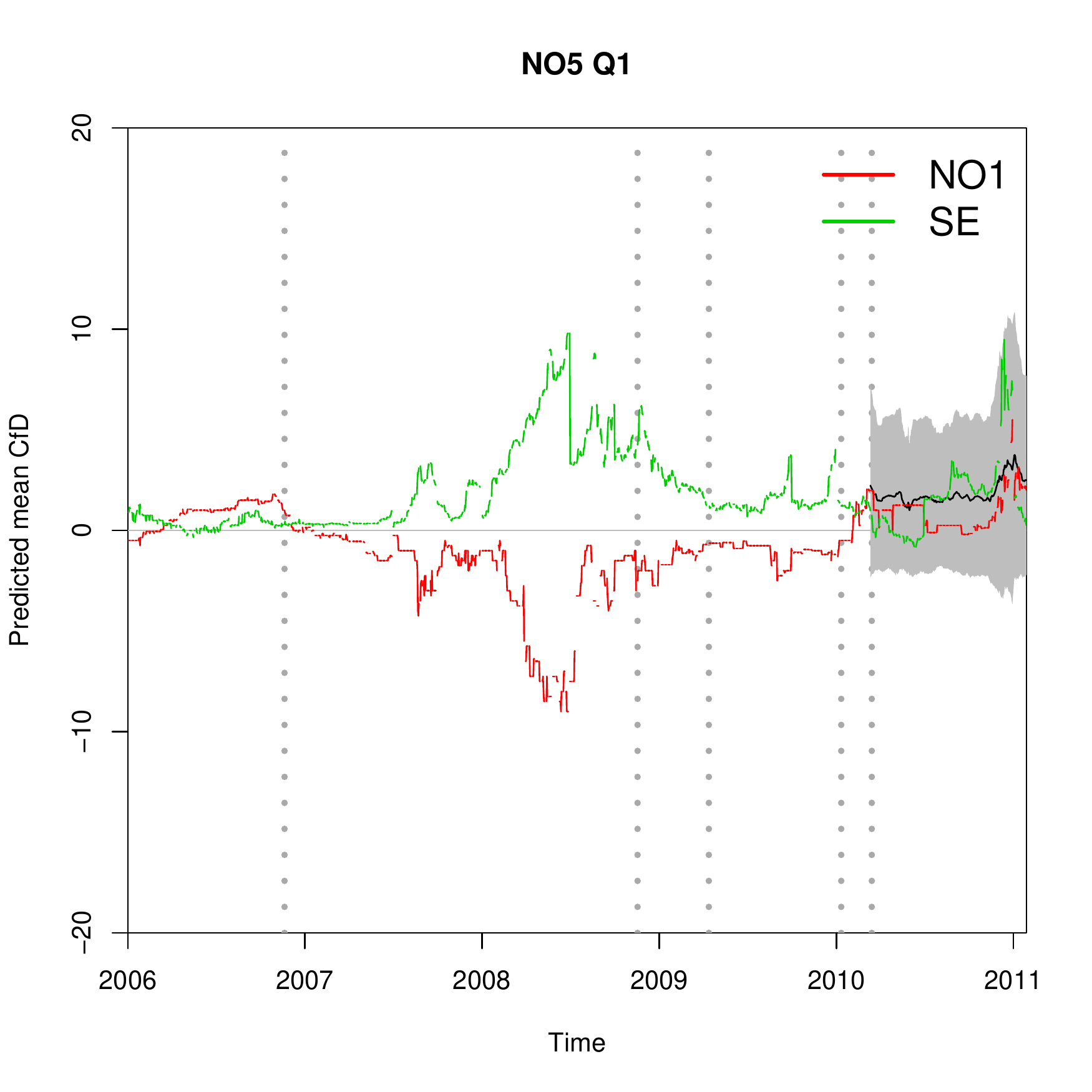}
  \end{minipage}
  \caption{Predicted CfDs for the Q1 horizon.  A a 95\% prediction
interval is shown in grey. Dashed vertical lines indicate dates when the area
definitions changed.}
  \label{fig:predQ1}
\end{figure}

\begin{figure}[htn]
  \centering
  \begin{minipage}{0.45\linewidth}
    \includegraphics[width=\linewidth]{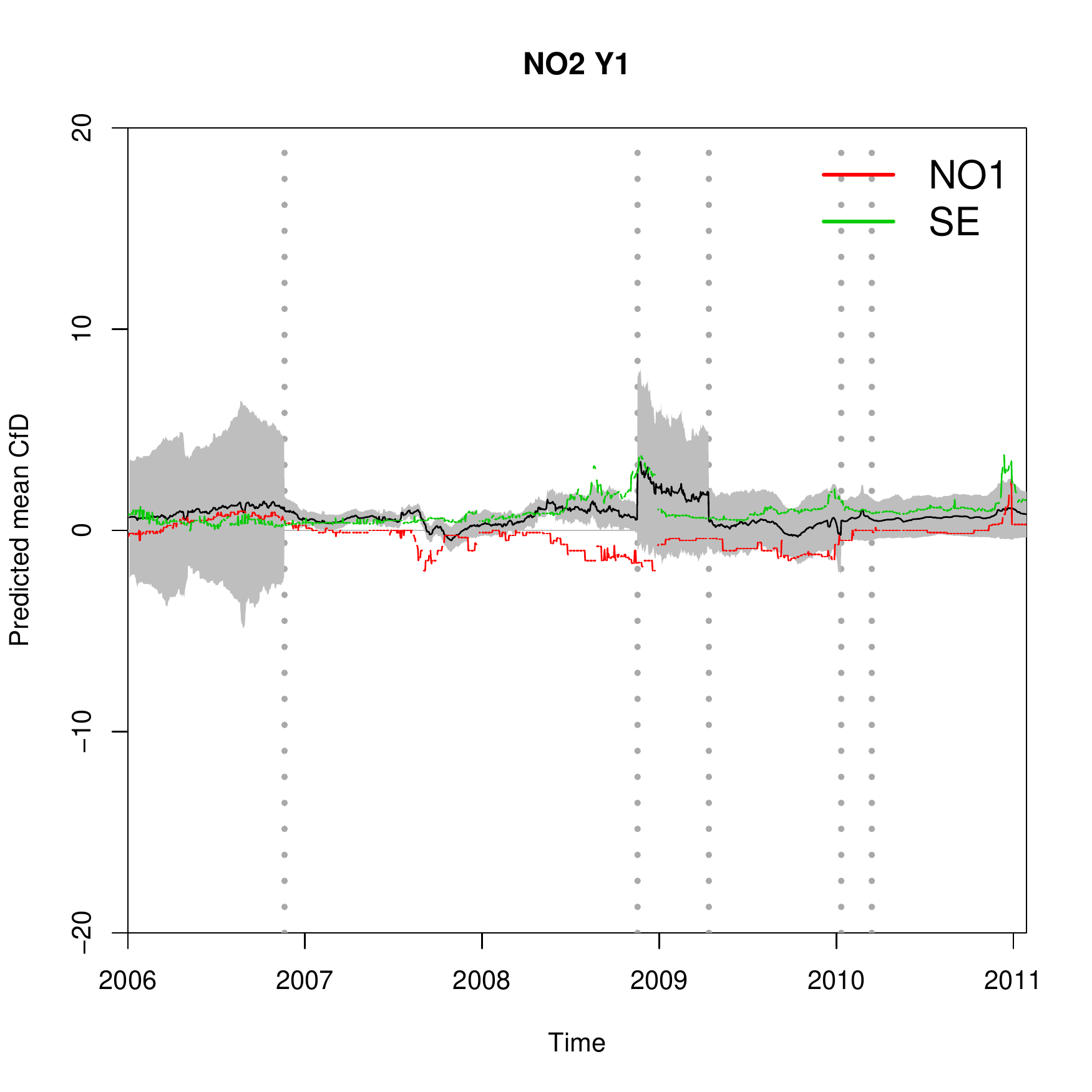}
    \includegraphics[width=\linewidth]{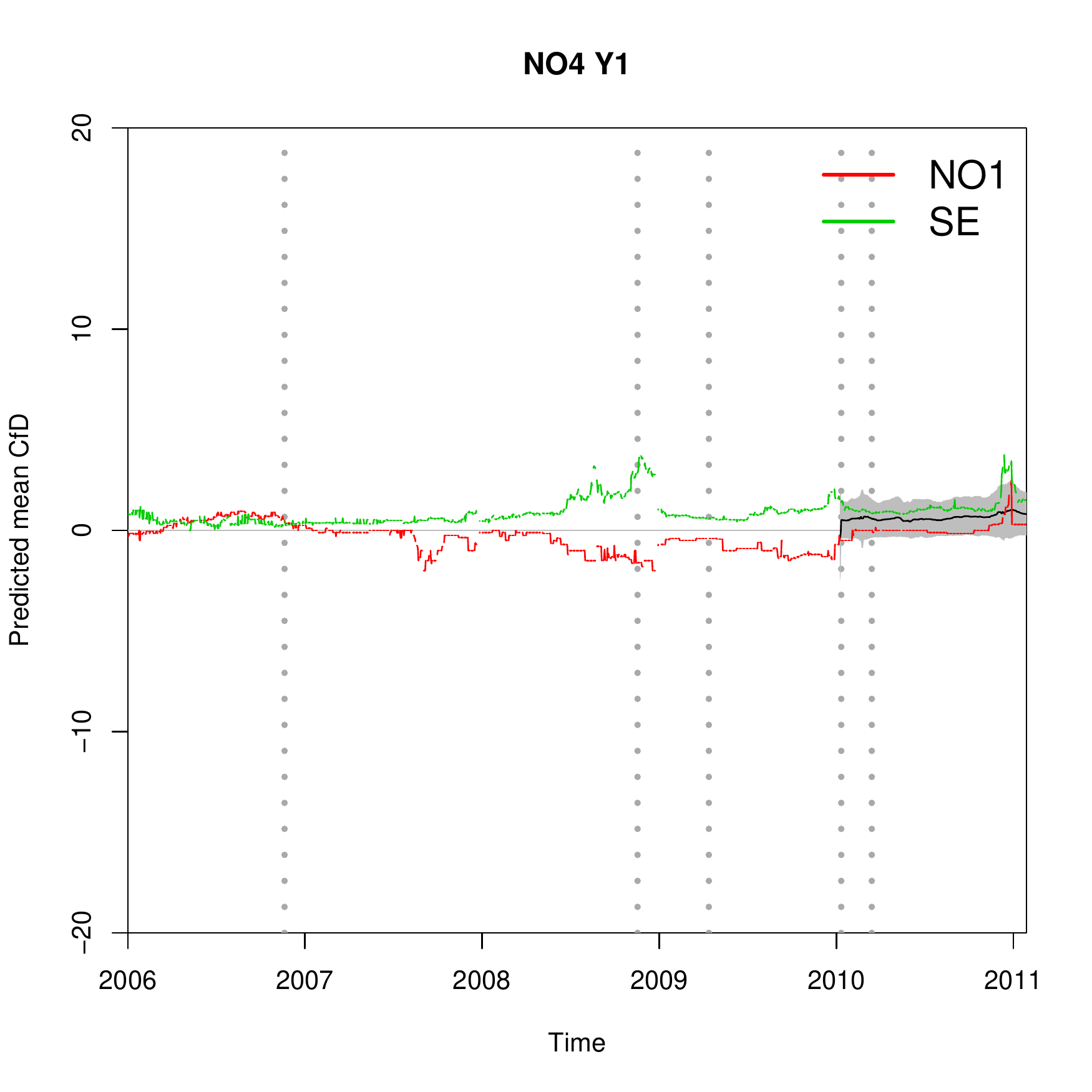}
  \end{minipage}
\begin{minipage}{0.45\linewidth}
    \includegraphics[width=\linewidth]{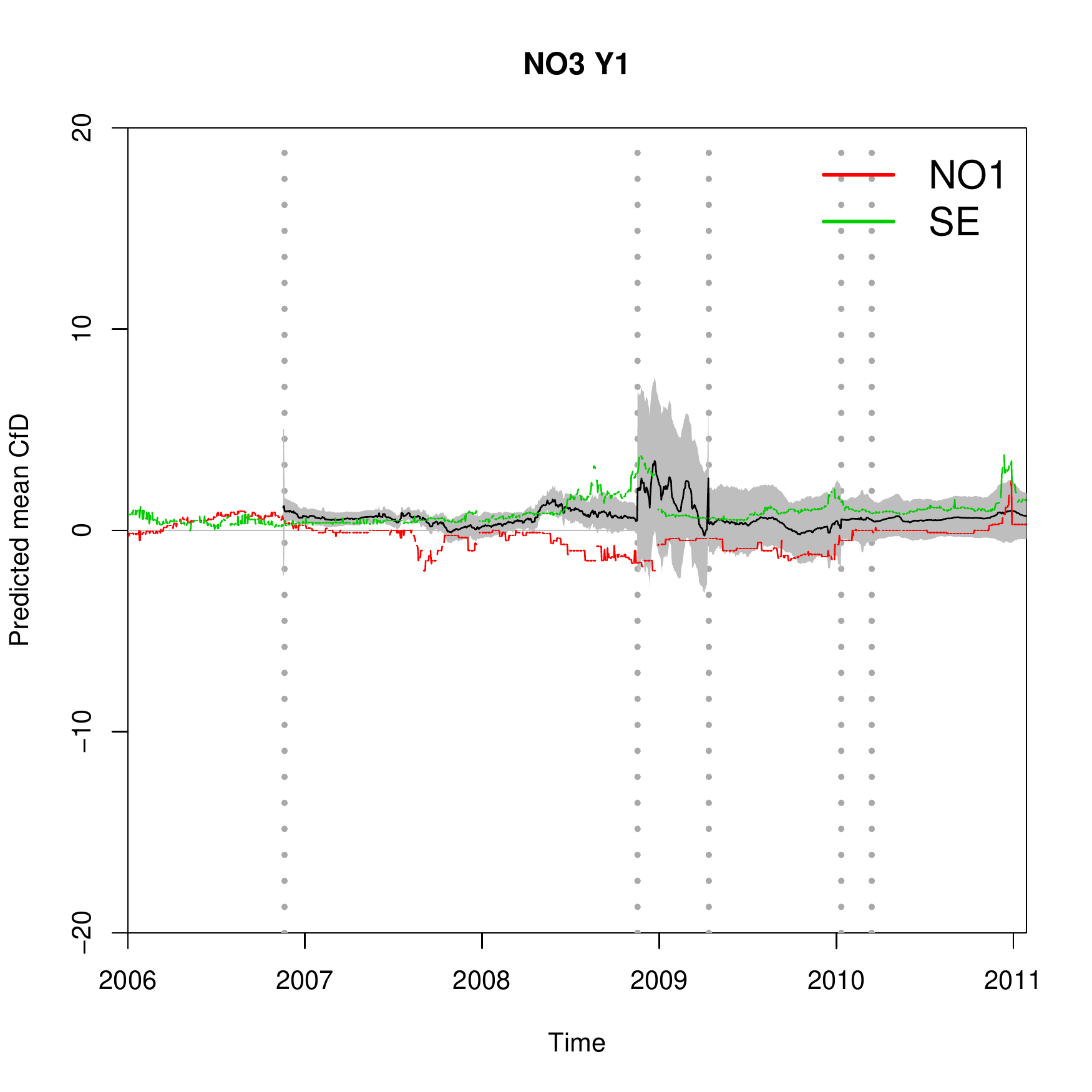}
    \includegraphics[width=\linewidth]{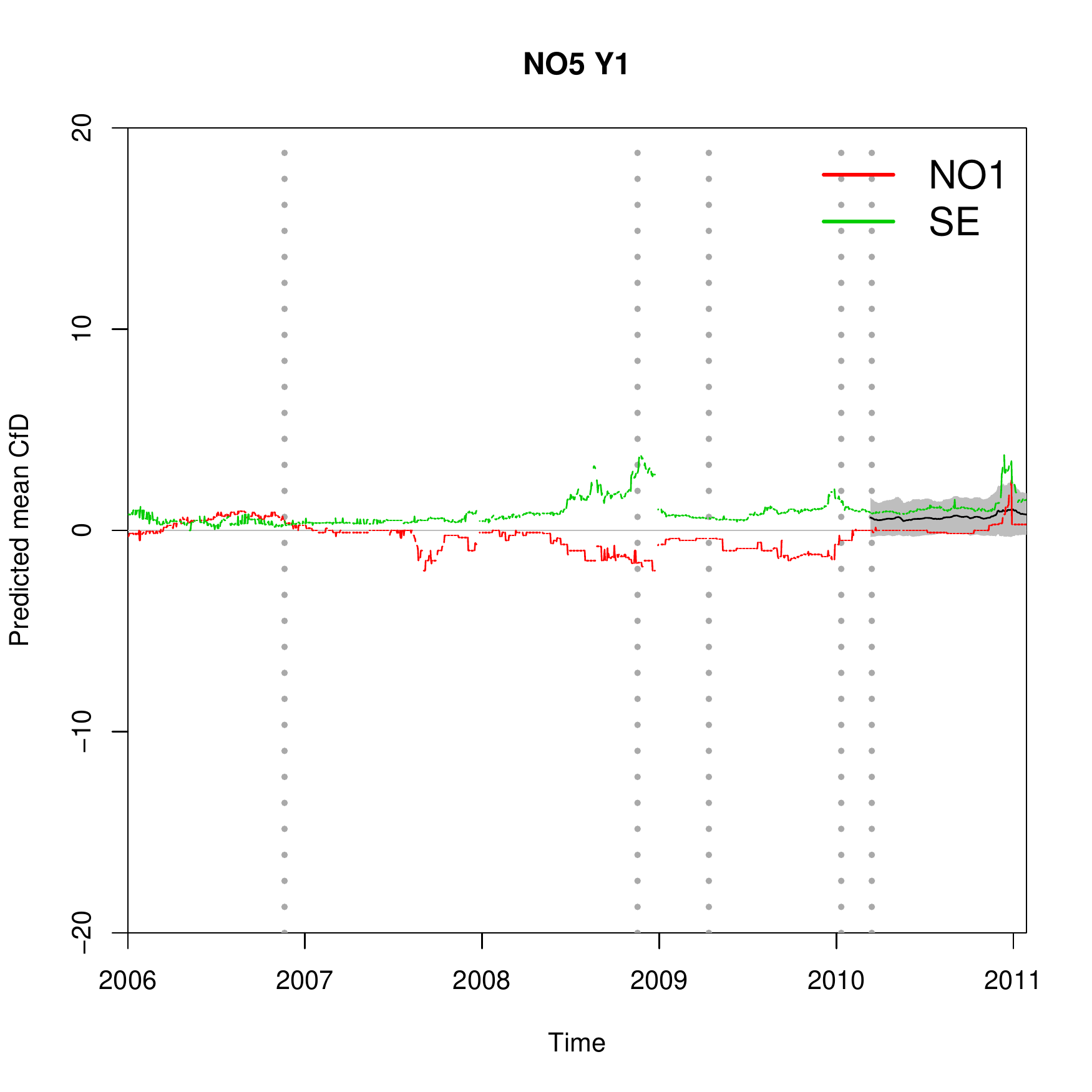}
  \end{minipage}
  \caption{Predicted CfDs for the Q1 horizon.  A a 95\% prediction
interval is shown in grey. Dashed vertical lines indicate dates when the area
definitions changed.}
  \label{fig:predY1}
\end{figure}

Figures~\ref{fig:predM1}--\ref{fig:predY1} show predicted, daily CfD prices
for horizons M1, Q1, and Y1, respectively, with a 95\% prediction
interval shown in grey, together with the observed NO1 and SE prices
for comparison.  For the M1 horizon, the predicted CfD prices for NO2 and NO3 are
quite similar, but there are some differences: For example, the NO3
price has a sudden
spike in the beginning of 2010, which is not seen for NO2. Both NO2
and NO3 seem quite similar to SE. The NO4 and
NO5 areas have only a rather short history, so it is not clear whether
they are more similar to NO1 or SE. For NO2 and NO3, we see that the
uncertainty varies a great deal between area definition periods ---
shorter periods have a larger uncertainty, since the amount of
available data is smaller. For the Q1 horizon, results look
similar. Finally, for the Y1 horizon, it is less clear whether the NO2
and NO3 prices are more similar to NO1 or SE, but they might still
seem slightly more like SE. The uncertainty is quite large at times,
particularly for NO2 in 2006. Estimated regression coefficients from the linear regression model in Equation~\eqref{eq:linmod} is shown in Table~\ref{tab:estregcoef}.

\begin{table}[ht]
\centering
\begin{tabular}{rrrrrr}
 \hline
\hline
\textbf{M1} & DK1 & DK2 & FI & NO1 & SE \\ 
  \hline
$\beta_{\SA}$ & 0.623 & 0.523 & 0.481 & 0.521 & 0.515 \\ 
  $\beta_{\SyS}$ & -0.151 & -0.050 & -0.439 & -0.449 & -0.460 \\ 
  $\beta_{\FW}$ & -0.325 & -0.304 & -0.003 & -0.097 & -0.008 \\
  $\beta_{\HY}$ & NA & NA & 1.200 & 1.061 & -0.885 \\ 
   \hline
  \hline
\textbf{Q1} & DK1 & DK2 & FI & NO1 & SE \\ 
  \hline
$\beta_{\SA}$ & 0.504 & 0.342 & 0.288 & 0.363 & 0.373 \\ 
  $\beta_{\SyS}$  & -0.163 & -0.152 & -0.189 & -0.390 & -0.227 \\
  $\beta_{\FW}$ & -0.174 & -0.002 & -0.062 & -0.005 & -0.102 \\ 
  $\beta_{\HY}$ &  NA & NA  & 2.216 & 0.906 & 0.723 \\ 
   \hline
 \hline
 \textbf{Y1} & DK1 & DK2 & FI & NO1 & SE \\ 
  \hline
$\beta_{\SA}$ & 0.076 & 0.114 & 0.059 & 0.029 & 0.023 \\ 
  $\beta_{\SyS}$  & 0.038 & 0.021 & -0.020 & -0.065 & -0.004 \\ 
  $\beta_{\FW}$ & 0.032 & 0.036 & -0.010 & 0.027 & 0.006 \\ 
  $\beta_{\HY}$ & NA  &  NA & 0.565 & -0.743 & -0.844 \\
   \hline
   
\end{tabular}
\caption{Estimated regression coefficients (rounded to three decimal places) from the linear regression model in Equation~\eqref{eq:linmod} for horizons M1, Q1, and Y1. Danish areas DK1 and DK2 have no regression coefficient for the reservoir level; since there is no hydropower production in Denmark, the reservoir level is left out from the linear model for the Danish areas.}
\label{tab:estregcoef}
\end{table}

\begin{figure}[ht]
  \centering
  \includegraphics[width=\linewidth]{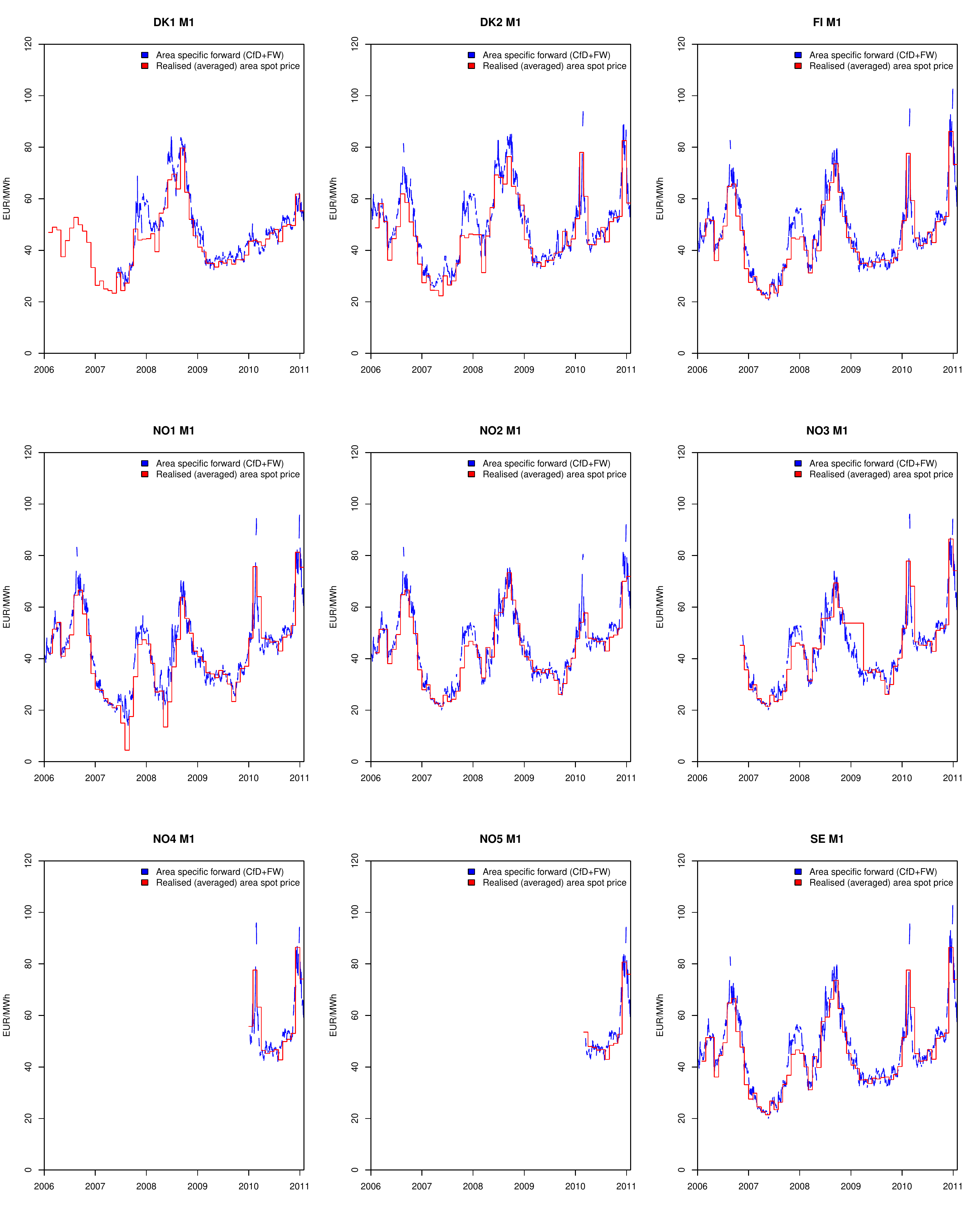}
  \caption{Predicted (NO2--NO5) and observed CfD + FW compared to the realised average
    area spot price for horizon M1.}
  \label{fig:rpm1}
\end{figure}

\begin{figure}[ht]
  \centering
  \includegraphics[width=\linewidth]{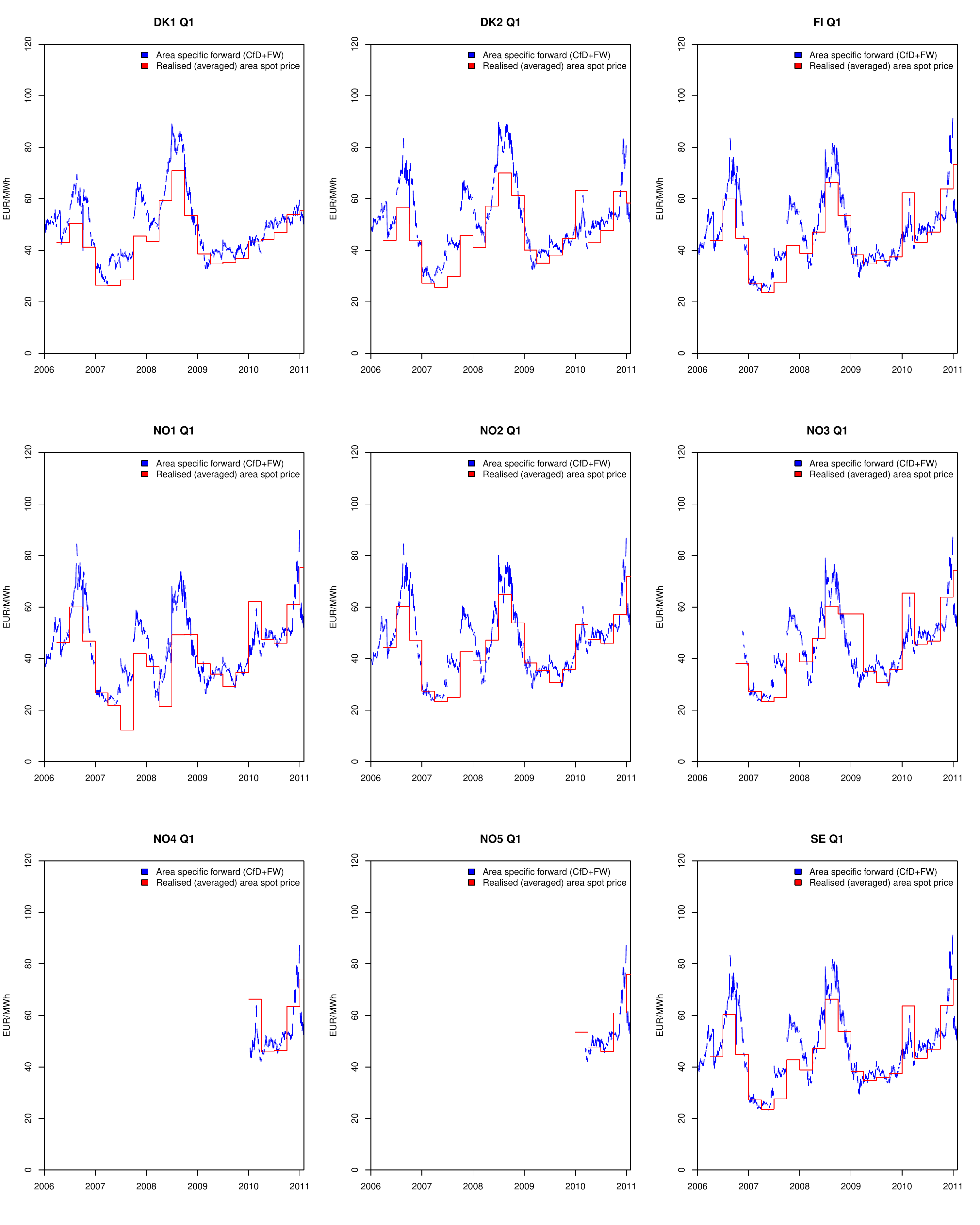}
  \caption{Predicted  (NO2--NO5) and observed CfD + FW compared to the realised average
    area spot price for horizon Q1.}
  \label{fig:rpq1}
\end{figure}

\begin{figure}[ht]
  \centering
  \includegraphics[width=\linewidth]{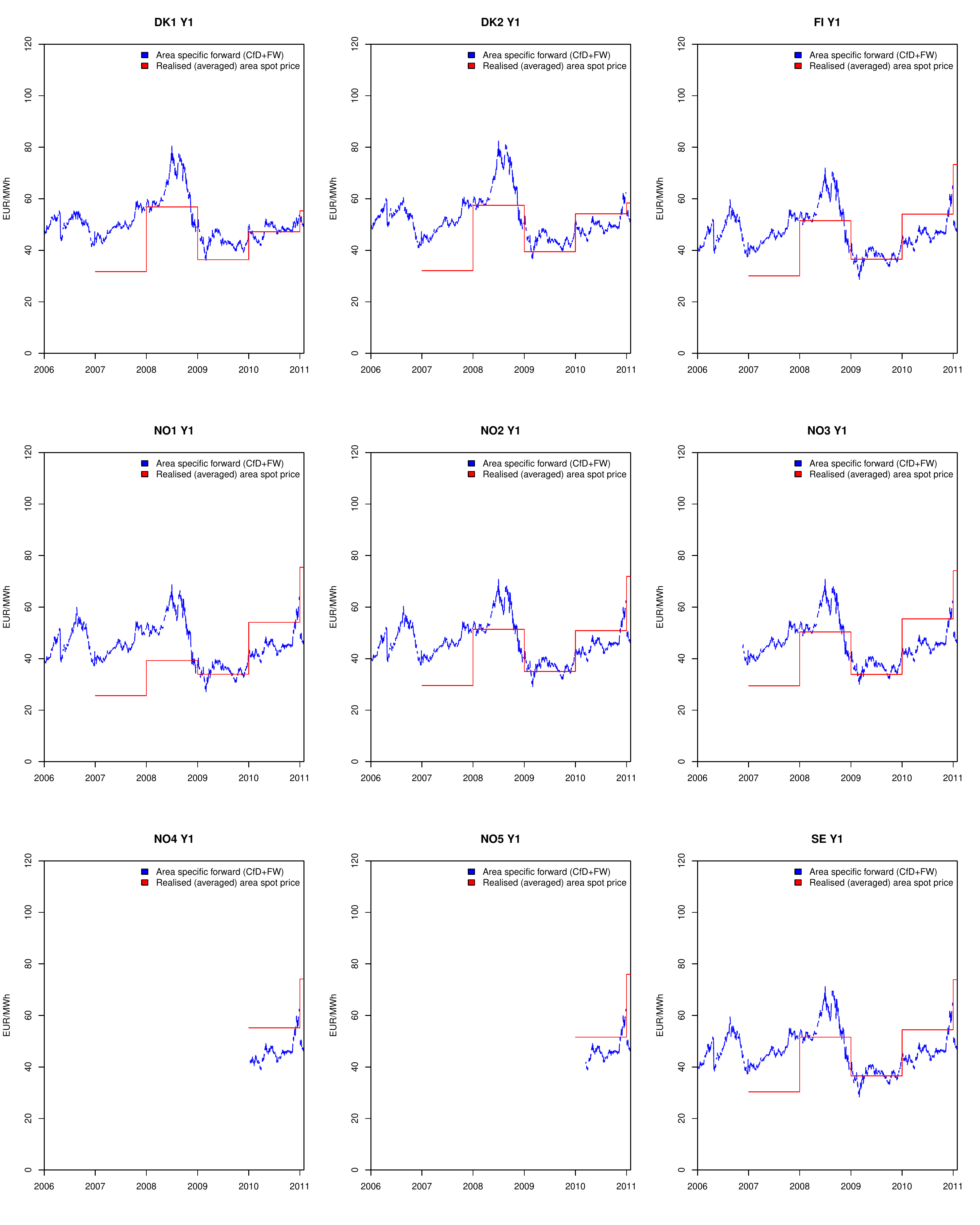}
  \caption{Predicted  (NO2--NO5) and observed CfD + FW compared to the realised average
    area spot price for horizon Y1.}
  \label{fig:rpy1}
\end{figure}

As explained in Section~\ref{sec:data}, for a given area, the sum of
the CfD and the corresponding system forward price may be interpreted
as an area-level forward price. Therefore, in the absence of risk
premia, each CfD $+$ FW sum should be similar to the realised 
area spot price, averaged over the contract period of the CfD. 
Figures~\ref{fig:rpm1}-\ref{fig:rpy1} show a comparison of observed (NO1, DK1, DK2, FI, SE) and predicted (NO2--NO5) CfD $+$ FW
and the realised averaged area spot price, for all price areas, and
for horizons M1, Q1 and Y1. Clearly, the correspondence is quite good between the observed and predicted CfDs, even for the longer-term horizons Q1 and Y1. 
For Y1, we only have a few years of observed data, and there is considerable uncertainty on how well the model performs for the longest horizons.  

\section{Discussion\label{sec:concl}}

We have asked and answered the hypothetical question: What would the
CfD market price have been, say in NO2, if it had been traded? We will not
know for certain, but we have presented a statistical approach to this
class of problems, which works almost without data. The approach can
be especially useful in markets 
where the price areas definitions change over time, as they have done
in the Nordic power market. Our method can be applied in OTC trading,
or to evaluate market prices if public trading of CfD products is introduced in
a (new) price area. 

We have suggested to use statistical elicitation for weighting
together the regression coefficients. Instead, we could have fitted one
common model for all price areas, and elicited on each regression
coefficient.  In our case, the price areas are too different for this to
work, but this might be different in other applications. However, the
expert would then need to have an opinion on the actual regression
coefficients, which we find unrealistic.

NASDAQ OMX Commodities's CfD products are traded for eight horizons
(one month ahead to three years ahead). We found that the effect of
each covariate is different for each horizon, and that a separate
model for each horizon is needed. Generally, there should be common
features for neighbouring horizons \citep{aas04, benth08}, especially
between those with delivery period of the same length, and some sort
of local shrinking could be applied \citep{hastie13}.

The other side of the coin is the risk premium problem, which
has been given more attention previously, especially for the system price
\citep{botterud10,marckhoff09,lucia11}. If there are no risk premiums,   
\begin{align*} 
\E_{t-k}[\mbox{SA}_{ta}] & = \ \mbox{FW}_t + \mbox{CfD}_{ta},
\end{align*}
or
\begin{align*} 
\E_{t-k}[\mbox{SS}_{ta}-\mbox{SA}_{ta}] & = \  \mbox{CfD}_{ta},
\end{align*}
where $\E_{t-k}$ is the expectation conditioned on all
relevant information available at time $t-k$. The risk premium can lie
in $\mbox{FW}_t$, $\mbox{CfD}_{ta}$ or both. Our method should work
equally well for risk premium estimation. A related problem would be
to build similar models for the expected area spot prices,
$\E_{t-k}[\mbox{SA}_{ta}]$, which could also benefit from our
approach.

The methods can be refined further, especially by thinking more on the
data process. The historical CfD prices are daily (five days a week)
closing prices, settled by NASDAQ OMX Commodities's procedure. This
means that the liquidity may vary between price 
areas, and some products (delivery periods) in some areas for some days have
not been traded at all \citep{frestad12}. 
 Typically, products
with a delivery period far ahead are traded less than other
products. Incorporating data on the traded volume as well as the
prices, including bid and ask prices, may enhance both the
hypothetical CfD prices, as well as our understanding of the CfD
products that are supposed to be traded.

\section*{Acknowledgments}
This work was funded by Statistics for Innovation, (sfi)$^2$, one of
the first 14 Norwegian centers for research-based innovation.

We thank R{\o}nnaug S{\ae}grov Mysterud, Stefan Erath, Arnoldo
Frigessi, Gianpaolo Scalia Tomba, David Hirst and Marion Haugen for useful
discussions, and Per Tore Jensen Lund of the Norwegian Water Resources
and Energy Directorate for supplying parts of the data.

\newpage

\bibliographystyle{apalike}  

\newpage
\listoffigures

\end{document}